\documentclass[conference,compsoc]{IEEEtran}

\usepackage{enumitem}
\usepackage{nccmath}

\ifCLASSOPTIONcompsoc
  \usepackage[nocompress]{cite}
\else
  \usepackage{cite}
\fi

\usepackage{subcaption}

\ifCLASSINFOpdf
   \usepackage[pdftex]{graphicx}
\else
   \usepackage[dvips]{graphicx}
\fi

\usepackage{amsmath}

\usepackage{algorithmic}

\ifCLASSOPTIONcompsoc
  \usepackage[caption=false,font=footnotesize,labelfont=sf,textfont=sf]{subfig}
\else
  \usepackage[caption=false,font=footnotesize]{subfig}
\fi

\usepackage{amsthm}
\usepackage{amssymb}
\usepackage{xcolor}

\begin{document}

\title{Using non-convex optimization in quantum process tomography: \\ Factored gradient descent is tough to beat}

\author{\IEEEauthorblockN{David A. Quiroga}
\IEEEauthorblockA{Department of Computer Science\\ Rice University\\ Houston, TX, USA\\
Email: daq3@rice.edu}
\and
\IEEEauthorblockN{Anastasios Kyrillidis}
\IEEEauthorblockA{Department of Computer Science\\ Rice University\\ Houston, TX, USA\\
Email: anastasios@rice.edu}
}

\maketitle

\begin{abstract}
We propose a non-convex optimization algorithm, based on the Burer-Monteiro (BM) factorization, for the quantum process tomography problem, in order to estimate a low-rank process matrix $\chi$ for near-unitary quantum gates.
In this work, we compare our approach against state of the art convex optimization approaches based on gradient descent. 
We use a reduced set of initial states and measurement operators that require $2 \cdot 8^n$ circuit settings, as well as $\mathcal{O}(4^n)$ measurements for an underdetermined setting. 
We find our algorithm converges faster and achieves higher fidelities than state of the art, both in terms of measurement settings, as well as in terms of noise tolerance, in the cases of depolarizing and Gaussian noise models.
\end{abstract}

\IEEEpeerreviewmaketitle

\section{Introduction}
\label{introduction}

Benchmarking quantum computers plays a vital role in the Near-Intermediate Scale Quantum (NISQ) era \cite{preskill2018quantum}. 
Using a NISQ computer, it is important to determine the accuracy of computations, in order to hope for solutions to real-world problems. 
Methods aim towards certifying the fidelity of quantum states \cite{altepeter2005photonic, aaronson2018shadow}, quantum processes \cite{altepeter2003ancilla, kunjummen2023shadow}, and application-based quantum results \cite{kardashin2020certified, eisert2020quantum}. 
Included in this type of methods are randomized benchmarking (RB) \cite{knill_randomized_2008}\cite{dankert_exact_2009}, cycle benchmarking (CB) \cite{erhard_characterizing_2019}, quantum state tomography (QST) protocols \cite{kyrillidis2017provable, gross2010quantum, banaszek1999maximum, paris2001maximum, rehacek2007diluted, gonccalves2012local, siah2013informationally, smolin2012efficient, Torlai2018Neural, Cramer2010Efficient, lanyon2017efficient, flammia2011direct, da2011practical, kalev2019validating}, 
gate-set tomography protocols  \cite{nielsen_gate_2021, blume-kohout_demonstration_2017}, and quantum process tomography (QPT) protocols \cite{mohseni2008quantum, jevzek2003quantum, kliesch2019guaranteed}, among others.

It is obvious from the above that QPT is somewhat less studied in the literature, compared to e.g., QST, mostly due to the computational requirements that come with such a study.
Yet, in this work, we focus on QPT:  
Put in simple words, QPT is the task of characterizing an unknown quantum process using measurement data \cite{mohseni2008quantum, jevzek2003quantum, kliesch2019guaranteed}. 
QPT has great importance when trying to determine the effects a quantum process has on a quantum computer, especially when not much information is known about the process, or when finding its fidelity towards a target quantum gate is required. 
Such a process is often considered a black box, where qubits are manipulated with imprecise control, inevitably inserting some degree of noise. 
By preparing and performing successive measurements on an unknown process $\mathcal{E}$, it is possible to estimate the process following specific constraints.

QPT was discussed by Chuang and Nielsen \cite{chuang_prescription_1997, nielsen_quantum_2010} as a theoretical procedure, and by Poyatos and Cirac \cite{poyatos_complete_1997} in an experimental setting.
In the latter, the setting uses QST, along with an inversion map as a final step, in order to estimate the process matrix. 
Since then, other procedures for QPT that focus on optimization objectives are proposed, such as the projected least-squares on the process matrix representation \cite{surawy-stepney_projected_2022} and the gradient-descent on the Kraus representation \cite{ahmed_gradient-descent_2022} versions of QPT; for more information, please refer to the Related Works section of this paper.

For the case of low-rank analysis, QPT methods that make use of compressed sensing \cite{baldwin_quantum_2014} achieve computational advantages in time and resources required. 
Harnessing optimization objectives in QPT \cite{kalev2015quantum} enables incorporating previous knowledge about the problem, such as the completely positive (CP) and trace-preserving (TP) conditions, in order to obtain a physical quantum process as an outcome.

Common factor and a critical aspect of the success of these methods is the scaling factor, as performing QPT even on more than one qubit is a resource-intensive task. 
Particularly, performing QPT on quantum stacks today --that offer out-of-the-box solutions-- require $12^n$ circuit configurations, with $n$ being the number of qubits included in the process, to achieve an accurate estimate \cite{pelaez_euler-rodrigues_2022}\cite{volya_state_2023}\cite{bowman_hardware-conscious_2023}. 
Thus, experiments on e.g., 3 qubits or more become easily infeasible. 
Beyond the number of measurements required, increasing the number of qubits has further implication from a computational point of view, where classical approaches rely on convex optimization solvers and computational-heavy projections onto positive semi-definite cones \cite{kalev2015quantum}.

In this paper, we exactly focus on the computational aspects of solving the QPT problem and follow the optimization-based path to propose a novel non-convex optimization method for QPT. 
Our algorithm is a factored gradient descent variant \cite{kyrillidis2017provable, photonics10020116} that exploits the positive-definiteness and low-rankness of a process matrix $\chi \in \mathbb{C}^{(2^n)^2 \times (2^n)^2}$. 
Our method utilizes input and measurement samples for full characterization, and test noise models with $\mathcal{O}(4^n r)$ measurement settings for an underdetermined scenario, in order to determine which kind of noise the algorithm is most resilient to on how many measurements. 
Here, $r$ is the rank of the process matrix $\chi$, such that $r \ll (2^n)^2$. 
We harness compressed sensing to approximate the low-rank process matrix $\chi$ to favor unitary processes with $r=1$.
Our findings and contributions in the paper are summarized below: \vspace{-0.1cm}
\begin{itemize}[leftmargin=*]
    \item We propose a novel non-convex optimization method for QPT that utilizes a low-rank assumption in order to approximate near-unitary processes. \textit{This is a nontrivial extension of non-convex methods for QST, where additional constraints need to be handled in QPT.}
    \item We test our algorithm on coherent and incoherent noise models on an underdetermined system in order to find noise resilience using a reduced set of initial state and measurement operators.
    \textit{Our findings highlight the superior performance of our methodology, as compared to state of the art.}
\end{itemize}

\color{black}

\section{Notation and QPT-related definitions}
{\label{sec:notation}}
A quantum state $|\psi\rangle \in \mathbb{C}^{2^n}$ is a $2^n$-dimensional complex vector, where $n$ is the number of qubits. 
Quantum states are usually represented by a density matrix:
\begin{equation*}
    \rho = |\psi\rangle\langle\psi| \in\mathbb{C}^{2^n\times2^n},
\end{equation*}
formed as an outer product of the vector state representation with itself. 
The off-diagonal elements of $\rho$ may provide information about errors that are characteristic of a mixed state and, thus, can contain more detailed information about the state. 
The types of errors that a quantum state can be subject to are, in general, coherent and incoherent, and are modeled as noise channels \cite{gutierrez_errors_2016}.

A unitary quantum process drives $|\psi\rangle$ towards a new state $|\psi'\rangle$, through $U|\psi\rangle \rightarrow |\psi'\rangle$.
Here, $U$ is a unitary matrix in $\mathbb{C}^{2^n \times 2^n}$.
When using density matrices, this transformation is equivalently expressed by $U\rho U^\dagger \rightarrow \rho'$, where $U^\dagger$ corresponds to the conjugate adjoint of $U$. 

The most general way of representing a quantum process is with a completely positive (CP) trace-preserving (TP) linear map $\mathcal{E}(\rho): \mathbb{C}^{2^n \times 2^n} \rightarrow \mathbb{C}^{2^n \times 2^n}$.
The Kraus representation of the linear map is written as: \vspace{-0.2cm}
\begin{equation} \label{eq:kraus}
    \mathcal{E}(\rho) = \sum_{i = 0}^{K-1} A_i \rho A_i^\dagger \\[-8pt] \nonumber ,
\end{equation}
with Kraus rank $K$ and $\{A_i: A_i \in \mathbb{C}^{2^n \times 2^n}, ~\forall i\}$ being the Kraus operators. 
When $A_i$ is a unitary matrix, then $K = 1$ and $A_0 = U$ \cite{baldwin_quantum_2014}. 
In fact, a linear map $\mathcal{E}(\cdot)$ is CP if it has a Kraus representation \cite{choi_completely_1975}. 

The Kraus operators can in turn be expressed using a fixed set of basis operators $\tilde{A}_m \in \mathbb{C}^{2^n \times 2^n}$: \vspace{-0.2cm}
\begin{equation} \label{eq:basis}
    A_i = \sum_m^b a_{im}\tilde{A}_m.
\end{equation}
where $a_{im} \in \mathbb{C}$ and $b$ refers to the number of basis operators the Kraus operators will be decomposed into.
One convenient choice of basis operators is \cite{chuang_prescription_1997, nielsen_quantum_2010}: \vspace{-0.2cm}
\begin{equation*}
    \tilde{A}_0 = I, ~\tilde{A}_1 = \sigma_x, ~\tilde{A}_2 = -i\sigma_y, ~\tilde{A}_3 = \sigma_z, \\[-8pt] \nonumber
\end{equation*}
where: \vspace{-0.2cm}
\begin{equation*}
    \sigma_x = \begin{bmatrix}
        0 & 1 \\
        1 & 0
    \end{bmatrix}, \sigma_y = \begin{bmatrix}
        0 & -i \\
        i & 0
    \end{bmatrix}, \sigma_z = \begin{bmatrix}
        1 & 0 \\
        0 & -1
    \end{bmatrix}. \\[-5pt] \nonumber
\end{equation*}
We can now define the process matrix representation, which is expressed as: \vspace{-0.2cm}
\begin{equation} \label{eq:chi}
    \mathcal{E}(\rho) = \sum_{n,m=1}^{b^2} \chi_{nm} \tilde{A}_m\rho\tilde{A}^\dagger_n, \\[-8pt] \nonumber
\end{equation}
where $\chi_{nm} = a_n a_m^\dagger$, $\chi \in \mathbb{C}^{b^2 \times b^2}$. $\chi$ thus contains the products of the coefficients of basis operators $\tilde{A}_n$ and $\tilde{A}_m$.

The number of fixed basis operators $b$
is found to be equivalent to $2^n$, such as for the Pauli basis, a special case of the Gell-Mann basis that is used for QPT \cite{siewert_orthogonal_2022}, where the operators correspond to a combination of $\{I, \sigma_x, \sigma_y, \sigma_z\}$ over $n$ qubits. 
We will use $b=2^n$ for the rest of the paper.

Let $\mathcal{H}_1$ and $\mathcal{H}_2$ represent the input and output Hilbert spaces of $\mathcal{E}$, and $\mathcal{B}(\mathcal{H}_i)$ represent the set of all bounded operators acting on $\mathcal{H}_i$. For a CPTP linear map $\mathcal{E}$ with a Kraus representation and a positive semi-definite matrix $A$, the positive (P) condition in CP holds if: 
\begin{equation}
A\succeq 0 \in \mathcal{B}(\mathcal{H}_1) \longrightarrow \mathcal{E}(A)\succeq 0 \in \mathcal{B}(\mathcal{H}_2).
\end{equation}

Conversely, the CP condition holds if for an auxilliary Hilbert space $\mathcal{H}_a$, the following holds:
\begin{equation} \label{eq:A_CP}
    A' \succeq 0\in \mathcal{B}(\mathcal{H}_1 \otimes \mathcal{H}_a) \longrightarrow (\mathcal{E}\otimes I)(A') \succeq 0 \in \mathcal{B}(\mathcal{H}_2 \otimes \mathcal{H}_a).
\end{equation}
That is, for a positive semi-definite matrix $A'$ belonging to the space formed by $\mathcal{H}_1\otimes\mathcal{H}_a$, transforming $A'$ through the linear map $\mathcal{E}\otimes I$ would also yield a positive semi-definite matrix in the space formed by $\mathcal{H}_2\otimes\mathcal{H}_a$. This previous condition expands the use of linear maps to systems where ancilla qubits are used, as $\mathcal{H}_a$ is precisely the space where the ancilla qubits act as an auxiliary system, and the $I$ on the right hand side of eq. (\ref{eq:A_CP}) acts on the ancilla qubits by leaving them idle. The TP condition states that for all $A \in \mathbb{C}^{2^n \times 2^n}$ such that $A\succeq 0$, the following equality holds: \vspace{-0.2cm}
\begin{equation} \label{eq:TP}
    \text{Tr}(\mathcal{E}(A)) = \text{Tr}(A). \\[-8pt] \nonumber
\end{equation}
A similar condition for noisy linear maps is trace non-increasing, where \vspace{-0.2cm}
\begin{equation} \label{eq:TP}
    \text{Tr}(\mathcal{E}(A)) \leq \text{Tr}(A). \\[-8pt] \nonumber
\end{equation}

\section{Factored Gradient Descent in Tomography}

In the next subsection, we briefly describe the use of non-convex methods in QST, before we delve into the main contribution of this work, the non-convex factored gradient descent (FGD) algorithm for QPT.

\subsection{\texttt{FGD} for QST}

Quantum state tomography (QST) is the task of characterizing a quantum state given a list of measurement frequencies. 
It was explored in \cite{altepeter_photonic_2005} and has been studied as a convex optimization problem in \cite{kalev_quantum_2015} --with a focus on compressed sensing-- along with exploiting low-rankness and formulating it as a non-convex problem in \cite{kyrillidis_provable_2017}. 
A common and simple way to express QST is by formulating it as a least-squares problem: \vspace{-0.2cm}
\begin{equation} \label{eq:QST}
\begin{split}
    \underset{\rho \in \mathbb{C}^{2^n \times 2^n}}{\min} \text{ } & \text{ } F(\rho) := \tfrac{1}{2} \|f-\mathcal{A}(\rho)\|^2_2 \quad 
    \text{s.t.} \text{ } \text{ } \rho \succeq 0 \text{, ~} \text{Tr}(\rho) \leq 1.
\end{split}
\end{equation}
Here, we define the sensing mechanism $\mathcal{A}: \mathbb{C}^{2^n \times 2^n \rightarrow \mathbb{R}^n}$ via the Born rule $f_i := (\mathcal{A}(\rho))_i = \tfrac{2^n}{\sqrt{m}} \text{Tr}(P_i \cdot \rho)$, for $i=1, \hdots, m$ and for $P_i$ being random Kronecker combinations of Pauli matrices, with appropriate dimensions. %

The work done in \cite{kyrillidis_provable_2017} reflects a non-convex approach of performing QST. %
The idea is based on this simple observation: 
convex methods require expensive computations through procedures such as Lanczos method and SVD, in order to retrieve a target matrix that is positive semi-definite, which is one of the constraints of quantum density matrices in \eqref{eq:QST}. 
This adds computational complexity of the order $\mathcal{O}((2^n)^3)$. 
This overhead --observe that this is repeated per iteration of the algorithm-- makes the QST problem impractical starting at a small $n$, leading to limited research into characterizing states of quantum devices.

The BM factorization is based on the fact that a PSD matrix $\rho \in \mathbb{C}^{2^n \times 2^n}$ can be expressed as the product $\rho=UU^\dagger$, $U \in \mathbb{C}^{2^n \times r}$, where $r$ represents the rank of $\rho$. 
The work in \cite{kyrillidis2017provable} suggested Projected Factored Gradient Descent (\texttt{ProjFGD}): instead of working on the $\rho$ density matrices, \texttt{ProjFGD} operates on the factors $U$, leading to computational savings (both in floating points operations and computational memory). 
In math, the $\text{Tr}(\rho)\leq 1$ constraint is transformed into the convex constraint $\|U\|_F^2 \leq 1$ to result in the following, now non-convex, objective:
\begin{equation} \label{eq:LS-QST}
\begin{split}
    \underset{U \in \mathbb{C}^{2^n \times r}}{\min} \text{ } & \text{ } F(UU^\dagger) := \tfrac{1}{2} \|f-\mathcal{A}(UU^\dagger)\|^2_2 \quad 
    \text{s.t.} \text{ } \text{ ~} ||U||_F^2 \leq 1.
\end{split}
\end{equation}

The full per-iteration update of $U$ is based on a form of projected gradient descent over the factors $U$, leading to following non-convex recursion:
\begin{equation*}
    U_{t+1} = \Pi_\mathcal{C} (U_t - \eta \nabla_\rho F(U_t U_t^\dagger) \cdot U_t),
\end{equation*}
for a step size $\eta$. 
It is important to note that, since optimization is performed on the factors $U$, it is guaranteed by construction that the resulting estimate $\rho_t := U_t U_t^\dagger$ is always a positive semi-definite matrix. 
I.e., one can avoid the computationally heavy  convex projections, by directly working on the factors $U$.
Finally, the authors' findings show that random initialization of $\rho$ is sufficient for convergence of \texttt{ProjFGD}, and the projection $\Pi_\mathcal{C}(\cdot)$ is often unnecessary, leading to the \texttt{FGD} variant:
\begin{equation*}
    U_{t+1} = U_t - \eta \nabla_\rho F(U_t U_t^\dagger) \cdot U_t,
\end{equation*}
The accelerated version of this work can be found in \cite{photonics10020116}.

\subsection{Non-convexity for QPT}

We will now explain how one can utilize Burer-Monteiro factorization for QPT, and what considerations must be made. 
From an optimization point of view, a version of \texttt{FGD} tailored to QPT boils down to optimizing for $\chi$ instead of $\rho$, and applying the factorization when $2^n \rightarrow d^2$, with $d=2^n$. 
A quantum process could also be estimated through measurements on the process matrix representation, where the target variable is actually $U$ from $\chi = UU^\dagger$ for a specific rank $r$. 
The added complexity comes from the dimensionality of $\chi$, where for an order of $\mathcal{O}(4^n)$ measurements required in the traditional QST setting, the setting of its QPT counterpart requires $\mathcal{O}(16^n)$ measurements.\footnote{ 
The large scaling difference QPT has over QST could in turn lead to greater benefits when performing compressed sensing  in QPT. In this scenario, by setting $r$, we would require only $\mathcal{O}(4^n r)$ measurements, and $\mathcal{O}(4^n)$ in the case of a unitary process.}

In order to define direct QPT as an optimization problem, we would first require to select an adequate representation of the CP linear map $\mathcal{E}$, which contains the variables to optimize over. 
One of the most common representations to use in direct QPT is the process matrix representation expressed in \eqref{eq:chi}.
Based on this representation, a formal optimization procedure over $\chi$ can be described as:
\begin{equation} \label{eq:LS}
\begin{split}
    \underset{\chi \in \mathbb{C}^{d^2 \times d^2}}{\min} \quad & F(\chi) := \tfrac{1}{2} \sum_{i,j} \big(f_{ij}-\mathcal{A}_{ij}(\chi)\big)^2 \\
    \text{s.t.} \quad & \sum_{n,m}\chi_{nm}\tilde{A}_m^\dagger \tilde{A}_n = \mathbb{I}, \\
    & \chi \succeq 0, \quad \chi = \chi^\dagger.
\end{split}
\end{equation}
which corresponds to the least-squares (LS) representation of the procedure \cite{baldwin_quantum_2014}, and optimizes over $\chi$ by using the $\ell_2$-norm distance between the frequencies of the measurement results and the estimated frequencies. 
The estimated $\chi$ can then be substituted into eq. (\ref{eq:chi}) in order to determine $\mathcal{E}(\cdot)$.

\medskip
\noindent \textbf{QPT conditions.}
These optimization formulations are based on convex semidefinite programs (SDPs), with convex CP \eqref{eq:A_CP} and TP \eqref{eq:TP} constraints. 
To elaborate a bit more the above \eqref{eq:LS}, these conditions are: $i)$ $\chi \succeq 0$ is the CP condition, and $ii)$ $\sum_{n,m}\chi_{nm}\tilde{A}_m^\dagger \tilde{A}_n = \mathbb{I}$ is the TP condition, with $\text{Tr}(\mathcal{E}(A)) = 1$. Conversely, for trace non-increasing quantum processes, $\text{Tr}(\mathcal{E}(A)) < 1$. It is also useful to note that for the case of a knowingly real-valued symmetric $\chi$ and a Gell-Mann or Pauli basis $\{\tilde{A}_n\}$, $\sum_{n,m}\chi_{nm}\tilde{A}_m^\dagger \tilde{A}_n = \text{Tr}(\chi)$ due to anticommutative properties of the matrices of such bases.

\medskip
\noindent \textbf{Sensing mechanism.}
For the sensing mechanism $\mathcal{A}: \mathbb{C}^{2^n \times 2^n} \rightarrow \mathbb{R}^{2^n}$, %
we have: $\mathcal{A}_{ij} = \text{Tr}(D_{ij}^\dagger \chi)$ where $D_{ijmn} = \text{Tr}(\rho_i^{in} \tilde{A}_n^\dagger E_j \tilde{A}_m)$ when $D_{ij} \in \mathbb{C}^{d^2 \times d^2}$ is represented in the $\{\tilde{A}_n\}$ basis. 
$\{\tilde{A}_n\}$ can be predefined as an orthonormal basis with $\tilde{A}_n \in \mathbb{C}^{2^n \times 2^n}$, and $E_j$ corresponds to the elements of an arbitrary positive operator-valued measure (POVM). 
$\rho_i^{in}$ is explained later in the text.

\medskip
\noindent \textbf{Initial states and measurement operators.}
QPT implementations utilize a set of input states $\{\rho_i^{in}\}$ that form a basis for representing arbitrary states, as well as a set of positive operator-valued measure (POVM) matrices $\{E_i\}$ that perform an informationally complete measurement. These bases are commonly implemented as rotations over the $(x, y, z)$ coordinates of a qubit represented as a Bloch sphere through the transformation $|\phi^{in}\rangle = G|0\rangle$, from which $\rho^{in} = |\phi^{in}\rangle\langle\phi^{in}|$. The set of generic states $|\phi^{in}\rangle$ typically used in QPT is: 
\begin{equation} \label{eq:generic_initial_states}
\begin{split}
    & |k\rangle, k=0, \hdots, d-1 \\
    & \frac{1}{\sqrt{2}} (|k\rangle + |n\rangle), k=0, \hdots, d-2, n = k+1, \hdots, d-1, \\
    & \frac{1}{\sqrt{2}} (|k\rangle + i|n\rangle), k=0, \hdots, d-2, n = k+1, \hdots, d-1,
\end{split}
\end{equation}
and can be implemented through the Pauli preparation basis $\{|0\rangle, |1\rangle, |+\rangle, |+i\rangle \}^{\otimes n}$ with gates $G = \{I, X, H, H\text{ } S\}^{\otimes n}$.\footnote{These states, however, are not the only selection that can be made. In general, we require the use of unitarily informationally complete (UIC) sets as input states in order to distinguish G. That is, a set $\{\rho_i^{in}\}$ that undergoes unitary evolution as $\rho_i^{out} = G\rho_i^{in}G^\dagger$ is a UIC if and only if it distinguishes $G$ from any other CPTP map \cite{baldwin_quantum_2014}.}
We use the generic states from eq. (\ref{eq:generic_initial_states}) for our experiments, as they are easy to implement on a physical quantum device and are used in out-of-the-box implementations. An absolute minimum of $2$ input states could be used, although reliably reproducing one of the two states is non-trivial. \cite{baldwin_quantum_2014}

For the measurement operators, however, a great reduction in the number of circuit settings can be made by selecting an appropriate POVM. POVM matrices are positive semi-definite Hermitian matrices that satisfy $\sum_{i} E_i = \mathbb{I}$, and are applied to a state $\rho$ in order to perform a measurement in the form of $\text{Tr}(E_i \rho)$.
In order to satisfy this requirement, we must define a POVM that completely characterizes an output state. The traditional setting utilizes the $\{\sigma_x, \sigma_y, \sigma_z \}^{\otimes n}$ basis that is implemented by applying quantum gates $M = \{I, H, SDG\text{ } H\}^{\otimes n}$ at the end of the circuit. Here we recover a reduced choice of POVM elements restated in a more specific context in \cite{baldwin_quantum_2014}, where $2d$ operators are chosen for a measurement that is informationally complete for pure states:
\begin{equation}
    \begin{split}
        & E_0 = a|0\rangle \langle 0| \\
        & E_n = b(1 + |0\rangle \langle m| + |m\rangle \langle 0|)\text{, } m=1,\hdots, d-1, \\
        & \tilde{E}_n = b[1 + i(|0\rangle \langle m| - |m\rangle \langle 0|)]\text{, } m=1,\hdots, d-1, \\
        & E_{2d} = 1 \Bigr[E_0 + \sum_{n=1}^{d-1}(E_n + \tilde{E}_n) \Bigr],
    \end{split}
\end{equation}
where $a$ and $b$ are selected such that $E_{2d} \succeq 0$. These POVM elements thus completely determine all pure states in a $d$-dimensional Hilbert space. 
Compared to the traditional case of requiring $O(12^n)$ circuits, we require only $O(8^n)$ circuits for our completely determined system. For mixed states, complete characterization of a unitary map requires a minimum of $d^2 - 1 +2d$ POVM elements. In particular, $d^4 - d^2$ elements are required to characterize a general completely positive trace-preserving map.

\medskip
\noindent \textbf{Handling constraints.} 
Handling constraints is not easy; we are first interested in performing experiments on an optimizer that solves an approximation to \eqref{eq:LS}, where the TP condition is handled in the objective function. 
To do so, we study the case where the TP condition is included as a regularizer into the objective function by means of:
\begin{align*}
    \sum_{n,m}\chi_{nm}\tilde{A}_m^\dagger \tilde{A}_n = \mathbb{I} &\Rightarrow \\
    \sum_{n,m}\chi_{nm}\tilde{A}_m^\dagger \tilde{A}_n - \mathbb{I} \precnapprox	 \varepsilon \mathbb{I} &\Rightarrow \\
    \left \|\sum_{n,m} \chi_{nm}\tilde{A}_m^\dagger \tilde{A}_n - \mathbb{I} \right\|_F^2 \leq \varepsilon
\end{align*}
where $\varepsilon$ corresponds to an error threshold, dictated by numerical efficiency. %
Thus, as far as $\varepsilon$ is small, we are close to satisfying the constraint.
Then, one obtains the new --approximate-- objective: \vspace{-0.2cm}
\begin{equation} \label{eq:LS-regularizer}
\begin{split}
    \underset{\chi \in \mathbb{C}^{d^2 \times d^2}}{\min} \text{ } & F(\chi) + \lambda \cdot \left \|\sum_{n,m} \chi_{nm}\tilde{A}_m^\dagger \tilde{A}_n - \mathbb{I} \right\|_F^2\\
    \text{s.t.} \text{ } & \chi \succeq 0 \text{, } \chi = \chi^\dagger,
\end{split}
\end{equation}
with $\lambda \in \mathbb{R}$ being the regularization parameter, that balances the importance between the two objectives: whether a solution $\chi$ should minimize the least-squares fidelity term, $F(\chi) := \tfrac{1}{2} \cdot \sum_{i,j} \big(f_{ij}-\mathcal{A}_{ij}(\chi)\big)^2$, or the TP-basis regularization term, $\left \|\sum_{n,m} \chi_{nm}\tilde{A}_m^\dagger \tilde{A}_n - \mathbb{I} \right\|_F^2$. 
For the rest of the discussion, we will define $H(\chi) := \left \|\sum_{n,m} \chi_{nm}\tilde{A}_m^\dagger \tilde{A}_n - \mathbb{I} \right\|_F^2$.

We now apply the Burer-Monteiro (BM) factorization on $\chi$ of eq. (\ref{eq:LS-regularizer}). 
Performing the BM factorization eliminates the CP PSD constraint $\chi \succeq 0$ through $\chi = UU^\dagger$, $U\in \mathbb{C}^{d^2\times r}$ for the case of a hermitian matrix $\chi$, and thus results in the following objective:
\begin{equation} \label{eq:LS-BM-regularizer}
\begin{split}
    \underset{U \in \mathbb{C}^{d^2 \times r}}{\min} \text{ } F(UU^\dagger) + 
    \lambda \cdot H(UU^\dagger),
\end{split}
\end{equation}
where:
\begin{align*}
    F(UU^\dagger) &:= \tfrac{1}{2} \cdot \sum_{i,j} \big(f_{ij}-\mathcal{A}_{ij}(UU^\dagger)\big)^2 \\[-19pt] \\
    H(UU^\dagger), &:= 
    \left\|\sum_{n,m}(UU^\dagger)_{nm}\tilde{A}_m^\dagger \tilde{A}_n - I\right\|_F^2.
\end{align*}
Now, \eqref{eq:LS-BM-regularizer} is in an unconstrained optimization form, where one can apply factored gradient descent, as in:

\begin{equation*}
    U_{t+1} = U_t - \eta \cdot \left(\nabla_\chi F(U_t U_t^\dagger) + \lambda \cdot \nabla_\chi H(U_t U_t^\dagger)\right)\cdot U_t.
\end{equation*}

Next, we describe how these gradients can be calculated to perform \texttt{FGD} on quantum process tomography.

\medskip
\noindent \textbf{Gradient calculations in non-convex objective \eqref{eq:LS-BM-regularizer}.}
Based on the discussion above, one needs to obtain exact descriptions of the gradient terms, $\nabla_\chi F(U_t U_t^\dagger)$ and $\nabla_\chi H(U_t U_t^\dagger)$.

The former is already contained in prior work on \texttt{FGD} for QST \cite{kyrillidis2017provable, photonics10020116} and it is equivalent to: 
\begin{align*}
    \nabla_\chi F(U_t U_t^\dagger) := \nabla F(\chi_t) \cdot U_t = -\mathcal{A}^\dagger\big(f-\mathcal{A}(\chi_t)\big) \cdot U_t,
\end{align*}
where the operator $\mathcal{A}^\dagger: \mathbb{R}^m \rightarrow \mathbb{C}^{d^2 \times d^2}$ is the adjoint of $\mathcal{A}$.

The term $\nabla_\chi H(U_t U_t^\dagger)$ requires some care.
Similarly to above, we have:
\begin{align*}
    \nabla_\chi H(U_t U_t^\dagger) := \nabla H(\chi_t) \cdot U_t.
\end{align*}
where for all $\alpha, \beta \in [d^2]$, the $(\alpha, \beta)$-th entry of the gradient term $\nabla H(\chi_t) \in \mathbb{C}^{d^2 \times d^2}$ satisfies:
\begin{small}
\begin{align} \label{eq:grad_h_indicated}
    \left[\nabla H(\chi_t)\right]_{\alpha, \beta} &= \frac{\partial \left(\text{Tr}\left(\left(\sum_{ij} \chi_{ij} B_{ij} - \mathbb{I}\right)^\dagger\left(\sum_{ij} \chi_{ij} B_{ij} - \mathbb{I}\right)\right)\right)}{\partial \chi_{\alpha \beta}} \nonumber \\ &= c_1 + c_2 + c_3.
\end{align}
\end{small}
where $B_{ij} := \tilde{A}_j^\dagger \tilde{A}_i$ and:
\begin{align*}
    c_1 &= 2\text{Tr}\left(B_{\alpha \beta}^\dagger \left(\sum_{ij} \left(\chi_t\right)_{ij} B_{ij}\right)\right),  \\
    c_2 &= -\text{Tr}(B_{\alpha \beta}^\dagger), \\
    c_3 &= -\text{Tr}(B_{\alpha \beta}).
\end{align*}
The complete expression is therefore:
\begin{align*}
    \left[\nabla H(\chi_t)\right]_{\alpha, \beta} &= 2 \text{Tr}\left(B_{\alpha \beta}^\dagger\left(\sum_{ij} \left(\chi_t\right)_{ij} B_{ij} - I\right)\right).
\end{align*}

\medskip
\noindent \textbf{Step size $\eta$ selection.}
We harness smoothness considerations on $\nabla_\chi F(U_t U_t^\dagger)$ and $\nabla_\chi H(U_t U_t^\dagger)$ in order to reach an adequate step size $\eta$. Take $\nabla f(\chi) = \nabla_\chi F(\chi) + \lambda \cdot \nabla_\chi H(\chi)$. By using Lipschitz continuity in the form of:
\begin{align*}
    \| \nabla f(\chi) - \nabla f(\zeta)\|_2 \leq L \| \chi - \zeta \|_2,
\end{align*}
we easily determine a loose upper bound for GD with
\begin{align*}
    L = \frac{4^n}{m} \left\| \sum_i D_i \|D_i^\dagger\|_F \right\|_F  + 2^{6n+1}.
\end{align*}
Nonetheless, adaptive versions of these algorithms such as adaptive gradient descent (\texttt{adaGD}) and adaptive factored gradient descent (\texttt{adaFGD}) provide a better estimate for $\eta$ and thus reduce the number of iterations required to reach convergence. For the adaptive algorithms, the step size $\eta$ is variable and can be set by means of 
\begin{align*}
    \eta_t \propto \frac{\| \mathcal{A}^\dagger (\mathcal{A}(U_t U_t^\dagger) - f) \|_2}{\| \mathcal{A}(U_t U_t^\dagger)\|_2},
\end{align*}
as stated in previous work \cite{kyrillidis2013matrix}. We will use \texttt{adaGD} and \texttt{adaFGD} with this step size for the remainder of our work.

\subsection{Error models}
Quantum processes are not fault-tolerant in the NISQ era. 
One of the goals of this work is to study \texttt{FGD} on QPT under various noise models and provide evidence that this methodology is more noise-resilient, as compared to classic convex optimization methods.
Our hypothesis is that the explicit inclusion of the low-rank constraint via the matrix factorization $\chi = UU^\dagger$ functions as a noise-cancelling regularization, compared to convex methods that do not explicitly use the prior knowledge that $\chi$ is of low-rank.

There are many types of noise sources that may affect a quantum process in any step of the tomography, namely the state preparation, measurement, and computation steps. These noise sources can mostly be separated into two types: coherent and incoherent \cite{feng_estimating_2016}. A general expression for a noisy quantum channel applied to a quantum state is $\mathcal{E}_a = \mathcal{E}_{err} \circ \mathcal{E}_t$, where $\mathcal{E}_t (\rho_{\text{init}}) = U_t\rho_{\text{init}}U_t^\dagger$ is the target quantum channel we are given to measure. 

In the measurement setting, a general Gaussian additive model can redefine a measurement as:
\begin{equation} \label{eq:gaussian-model}
    \mathcal{A}_{ij} = \text{Tr}(D_{ij}^\dagger \chi) + \xi \epsilon_{ij},
\end{equation}
where $\epsilon_{ij}$ is a Gaussian random variable with $0$ mean and $1$ variance, and $\xi = [0,1]$ is our noise parameter. We continue to use $\xi$ in the formulations of all the error models.

Coherent errors are represented by systematic rotations of a state along a certain axis and have cumulative effects on the fidelity of a state. In fact, the most representative noise in a quantum device will be coherent \cite{bravyi_correcting_2018} as it accumulates quadratically. These errors are caused by control noise, external fields, qubit-qubit interactions and cross-talk \cite{greenbaum_modeling_2018, quiroga_discriminating_2021}. This type of errors can be represented as:
\begin{equation} \label{eq:coherent-model}
\mathcal{E}_{err}(\rho) = U_{coh}\rho U_{coh}^\dagger
\end{equation}
with $U_{coh} = e^{i\xi H}$. $H$ is a Hermitian matrix used to create over-rotation, and which we randomly generate.

Incoherent errors, on the other hand, are defined as statistical errors that accumulate linearly and couple a quantum system with the environment. In addition to having less impact than coherent errors, incoherent errors can be modeled as depolarizing noise and are thus easier to handle \cite{pascuzzi_computationally_2022}. The depolarizing noise model for incoherent errors is:
\begin{equation} \label{eq:depolarizing-model}
    \mathcal{E}_{err}(\rho) = (1-\xi)\rho + \tfrac{\xi}{d} \mathbb{I},
\end{equation}
where $\xi$ is the depolarizing strength.

Work done in \cite{baldwin_quantum_2014} uses a different model for incoherent errors, where random Kraus operators $\{A_i\}$ are generated using the Haar measure, and applied through the Kraus representation of eq. (\ref{eq:kraus}) paired with a noise parameter $\xi$ to produce the error model: \vspace{-0.2cm}
\begin{equation} \label{eq:incoherent-model}
    \mathcal{E}_{err}(\rho) = (1-\xi)\rho + \xi \sum_{i = 0}^{d^2 - 1} A_i \rho A_i^\dagger. \\[-6pt]
\end{equation}
This error model can reproduce the depolarizing model as well as models associated to bit flips, phase damping, and stochastic Pauli noise, with this last noise being able to represent the previous. Due to the flexibility of this model, we will consider it as a general incoherent noise model. 

\section{Related Work}
{\label{sec:QPT}}

\noindent \textbf{Standard Quantum Process Tomography (SQPT).} 
SQPT consists of recreating a quantum process $\mathcal{E}(\cdot)$ by $i)$ performing QST on a set of density matrices that represent a basis for an input state $\rho_{in}$, and $ii)$ selecting a set of basis operators $\{\tilde{A}\}$ from eq. (\ref{eq:basis}) that determine a $2^n \times 2^n$ matrix, from which a final inversion step allows the characterization of the parameters of $\mathcal{E}(\cdot)$. 

A general optimization procedure for calculating $\mathcal{E}(\rho_k)$, for a specific $\rho_k$, through QST is by formulating the least-squares optimization problem: \vspace{-0.2cm}
\begin{equation} \label{eq:SQPT}
\begin{split}
    \underset{\rho'_k \in \mathbb{C}^{2^n \times 2^n}}{\min} \text{ } & \frac{1}{2} \|f-\mathcal{A}(\rho'_k)\|^2_2 \\
    \text{s.t.} \text{ } & \rho'_k \succeq 0 \text{, } \text{tr}(\rho'_k) \leq 1, \rho'_k = \mathcal{E}(\rho_k)
\end{split}
\end{equation}
where $\rho'_k$ is a positive semi-definite and trace preserving (TP) matrix.
SQPT requires $d^2 = 4^n$ linearly independent $\rho_{in}$ inputs for which the output state $\mathcal{E}(\rho_{in})$ is determined through QST \cite{mohseni_quantum_2008}. 
Afterwards, we can employ eq. (\ref{eq:chi}) to construct a complex-valued matrix that would allow an appropriate matrix inversion. %

\medskip
\noindent \textbf{Ancilla-assisted Process Tomography (AAPT).} AAPT can be represented as a QST problem where a process $\mathcal{E}$ is characterized through the estimation of a state $\rho_\mathcal{E}$ and the use of ancilla qubits that may also capture information about the process. 
There exists two variations of AAPT where different types of measurements are applied, namely joint separable measurements and mutually unbiased bases measurements \cite{mohseni_quantum_2008}.

\medskip
\noindent \textbf{Direct Quantum Process Tomography.}
One of the most common representations to use in direct QPT is the process matrix representation expressed in \eqref{eq:chi}.
Based on this representation, a formal optimization procedure over $\chi$ can be described as: \vspace{-0.2cm}
\begin{equation}
\begin{split}
    \underset{\chi \in \mathbb{C}^{d^2 \times d^2}}{\min} \text{ } & \sum_{i,j} \left(f_{ij}-\mathcal{A}_{ij}(\chi)\right)^2 \\
    \text{s.t.} \text{ } & \sum_{n,m}\chi_{nm}\tilde{A}_m^\dagger \tilde{A}_n = 1, ~~\chi \succeq 0 \text{, } \chi = \chi^\dagger
\end{split}
\end{equation}
which corresponds to the least-squares (LS) representation of the procedure \cite{baldwin_quantum_2014}, and optimizes over $\chi$ by using the $\ell_2$-norm distance between the frequencies of the measurement results and the estimated frequencies. 
The estimated $\chi$ can then be substituted into eq. (\ref{eq:chi}) in order to determine $\mathcal{E}(\cdot)$.

Within direct QPT methods, compressed sensing reliably estimates a process matrix through the use of less measurements, by exploiting previous knowledge about the reconstructed matrix. 
This concept was first introduced in \cite{1614066} 
for classical signal recovery and later adapted to QPT \cite{kosut_quantum_2009, shabani_efficient_2011} and QST \cite{gross_quantum_2010, liu_universal_2011, flammia_quantum_2012}. 

We review the $\ell_1$-norm CS (CS$_{\ell_1}$) and the trace-norm CS (CS$_{\text{tr}}$) estimators, introduced and explained in \cite{baldwin_quantum_2014}.
When the $\chi$ matrix is close to a sparse matrix, it can be retrieved more efficiently by only minimizing its $\ell_1$-norm \cite{kosut_quantum_2009} and setting its previous loss function as another constraint with a threshold $\epsilon$, based on an estimation of the noise sources affecting the measurements. 
Given an orthogonal basis $\{V_n\}$ that includes the target unitary matrix $V_0 = U$, 
the CS$_{\ell_1}$ estimator may then be defined as:
\begin{equation} \label{eq:CS-l1}
\begin{split}
    \underset{\chi \in \mathbb{C}^{d^2 \times d^2}}{\min} \text{ } & \|\chi\|_1 \\
    \text{s.t.} \text{ } & \sum_{i,j} \left(f_{ij} - \mathcal{A}_{ij}( \chi)\right)^2 \leq \epsilon \\
    & \sum_{n,m} \chi_{nm}V_m^\dagger V_n = \mathbb{I}, ~~\chi = \chi^\dagger\text{, } \chi \succeq 0
\end{split}
\end{equation}
On the other hand, the CS$_{\text{Tr}}$ estimator is based on the assumption that $\chi$ is close to a low-rank matrix. 
In order to make use of low-rankness, we can minimize the nuclear norm/trace, as was shown in \cite{candes_stable_2006}. 
The authors of this method \cite{baldwin_quantum_2014} do so by taking an operator basis of traceless Hermitian matrices in order to maintain the maximal number of constraint equations, dropping only the equation relevant to the trace of $\chi$. 
The CS$_{\text{Tr}}$ estimator is thus formulated as:
\begin{equation} \label{eq:CS-Tr}
\begin{split}
    \underset{\chi \in \mathbb{C}^{d^2 \times d^2}}{\min} \text{ } & \text{Tr}(\chi) \\
    \text{s.t.} \text{ } & \sum_{i,j} (f_{ij} - \mathcal{A}_{ij}( \chi))^2 \leq \epsilon \\
    & \sum_{n,m \neq 1} \chi_{nm}\tilde{A}_m^\dagger \tilde{A}_n = 0, ~~\chi = \chi^\dagger\text{, } \chi \succeq 0 \\[-20pt] \nonumber
\end{split}
\end{equation}

\medskip
\noindent \textbf{Projected least-squares.}
Based on the projected least-squares state tomography method proposed in \cite{guta_fast_2018} and later extended to QPT in \cite{surawy-stepney_projected_2022}, the projected least-squares (PLS) QPT estimator implies the use of the LS estimator without constraints and includes a projection step onto the set of CPTP matrices. This method differs from CS in the sense that, given enough measurements, one can find a closed-form solution to the least-squares problem:
\begin{align*}
    \underset{\chi \in \mathbb{C}^{d^2 \times d^2}}{\min} \text{ }  \sum_{i,j} (f_{ij}-\mathcal{A}_{ij}(\chi))^2 &= \underset{\chi \in \mathbb{C}^{d^2 \times d^2}}{\min} \text{ } \|f-\mathcal{A}(\chi)\|_2 \\ &=
    (\mathcal{A}^\dagger \mathcal{A})^{-1}\mathcal{A}^\dagger(f)
\end{align*}
where $\mathcal{A}^\dagger$ is the adjoint of $\mathcal{A}$. Different techniques for projecting the estimated $\chi$ onto the CPTP set have been used, including projected gradient descent in \cite{knee_quantum_2018} and both Dykstra's algorithm \cite{dykstra_algorithm_1983} and the hyperplane intersection projection (HIP) algorithm in \cite{surawy-stepney_projected_2022}.

\medskip
\noindent \textbf{Gradient-descent quantum process tomography.}
QPT can also be performed by learning the Kraus representation of a process in the case of the gradient-descent quantum process tomography (GD-QPT) \cite{ahmed_gradient-descent_2022}. This provides an advantage over learning the complete set of parameters for the process representation using a Choi matrix, as only a fixed amount of Kraus operators need to be estimated. Despite the Choi matrix having a rank $r = 4^n$ when learning all its parameters, most real-world processes are low-rank and near-unitary with $r\ll 4^n$ while $r=1$ for the case of unitary processes. Conversely, the rank $r=k$ for a minimum fixed number of Kraus operators $k$ allows for a low-rank reconstruction of the process.

GD-QPT solves some of the limitations present in both CS and PLS QPT methods. GD-QPT can be used when not all the measurements are available due to low Kraus ranks like in CS, and may scale to larger problems as well alike PLS. GD-QPT is also less computationally expensive than CS and PLS, as the most expensive step in this procedure is the retraction where small matrices of dimensions $k2^n \times 2^n$ with Kraus rank $k\ll 4^n$ are inverted. On the other hand, PLS performs eigendecomposition of a Choi matrix of dimension $4^n \times 4^n$ resulting in qubic complexity.

\section{Results}

We first generate the measurements $f$ based on a set of initial probe states $\{\rho_i^{in}\}$, Gell-Mann basis matrices $\{\tilde{A}_n\}$, and measurement POVMs $\{E_j\}$ that construct $D_{ij}$ as explained in our setting. Following a mechanism from matrix sensing, our values for $f_{ij}$ are then generated by the model $f_{ij} = \mathcal{A}_{ij} (\chi^*) = \text{Tr}(D^\dagger_{ij} \chi^*)$, with $\chi^*$ being a rank-$r$ matrix. For noisy experiments, we include  errors through the models explained in the previous section. The approach we use to choose a valid $\chi^*$ is tailored to a rank-$1$ initialization. We construct a random Haar unitary matrix $H$, from which we define the linear equation $[\tilde{A}_{ij}] x = [H_{ij}]$, $x \in \mathbb{C}^{d^2 \times 1}$ and solve for $x$. $\chi^*$ can then be created by means of $\chi^* = xx^\dagger$.

The initialization of $\chi_0$ was done through a random complex matrix $M$ with one random real matrix $M_r$ and one random imaginary matrix $M_i$, both following a uniform distribution, such that $M = M_r + M_i$. %

\begin{figure}[t!]
    \centering
    \includegraphics[width=0.5\textwidth]{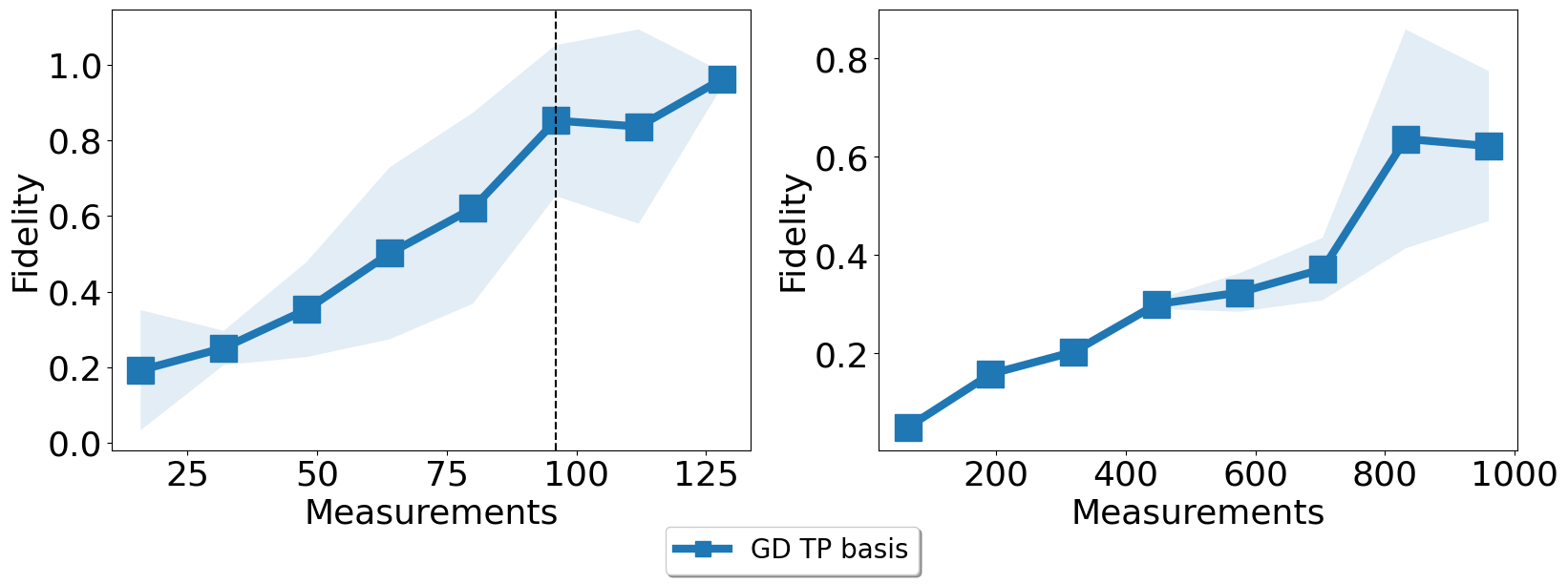}
    \caption{Number of noiseless measurements against fidelity using gradient descent for $2$ and $3$ qubits, respectively. Left plot is $10$k iterations and $10$ runs for 2 qubits, and right plot is $15$k iterations and $2$ runs for 3 qubits. The vertical line in the 2 qubit case shows the number of measurements we will take for the underdetermined experiment setting.} \vspace{-0.6cm}
    \label{fig:measurements_fidelity_GD}
\end{figure}

\medskip
\noindent \textbf{Fidelity scaling for number of measurements.}
We first set the number of measurements suitable for comparing algorithms in the case of an underdetemined system. To do this, we take the order of magnitude for the number of measurements required in CS for a rank-$r$ matrix $U$, $\mathcal{O}(r2^d)$, and define $m = Cr2^d$ with a constant $C$, as the real number of measurements to be taken. We use $C_2 = \{1, 2, \hdots, 8\}$ for the 2 qubit case and $C_3 = 2C_2$ for the 3 qubit case. We set $r=1$ as we're optimizing for a rank-$1$ matrix $\chi$. We then run the adaptive GD algorithm for different values of $C$ and take the fidelity of each scenario. The results are shown in Figure \ref{fig:measurements_fidelity_GD} for 2 qubits and 3 qubits. We set $\approx 0.8$ fidelity and $C_2=6$ as the baseline for all other undetermined experiments. After setting a $C$ to use in the undetermined settings, we run two sets of experiments, one with full measurements ($2r2^d$) and another with $C2^d$ measurements, and compare each noise model in order to find which algorithm performs best.

\begin{figure}[h]
    \centering
    \includegraphics[width=0.5\textwidth]{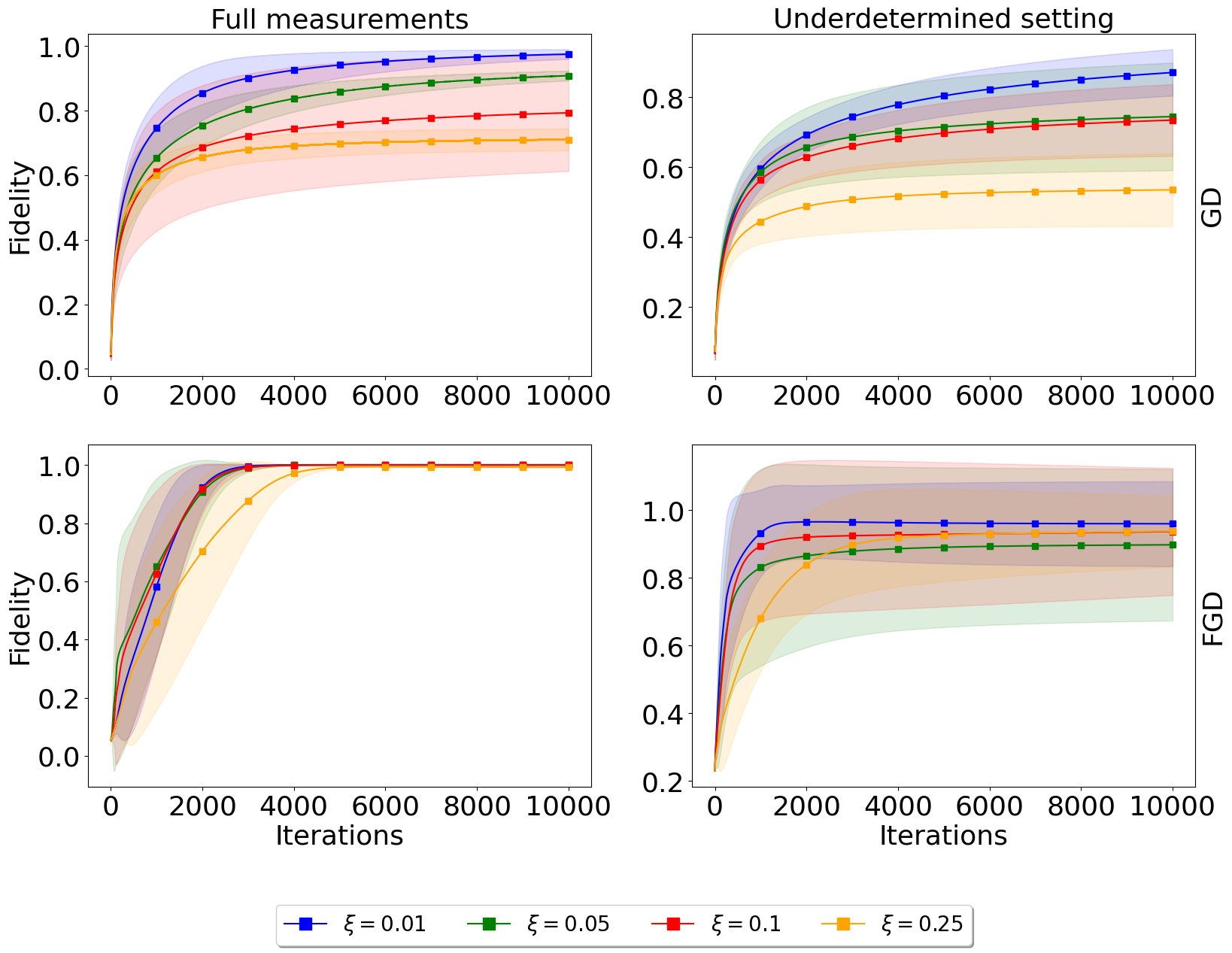}
    \caption{2 qubit depolarizing noise. The top and the bottom figures show results from the \texttt{adaGD} and the \texttt{adaFGD} algorithms, respectively. The left and right figures show results on 128 measurements and 96 measurements, respectively.}
    \label{fig:2_qubit_depolarizing_fidelity_paper}
\end{figure}

\medskip
\noindent \textbf{Depolarizing noise.} \texttt{FGD} reaches convergence to the optimal $\chi^*$ regardless of depolarizing noise in the full measurement setting of Figure \ref{fig:2_qubit_depolarizing_fidelity_paper}, while GD is more susceptible to it and converges to a suboptimal $\hat{\chi}$. The same can be said for the case of $C_2=6$, $m=96$, where we observe evidence towards \texttt{FGD} being less susceptible to depolarizing noise, although with increased variance. \texttt{FGD} showed great improvement for all depolarizing strengths $\xi$ tested on both measurement settings.

\begin{figure}[h]
    \centering
    \includegraphics[width=0.5\textwidth]{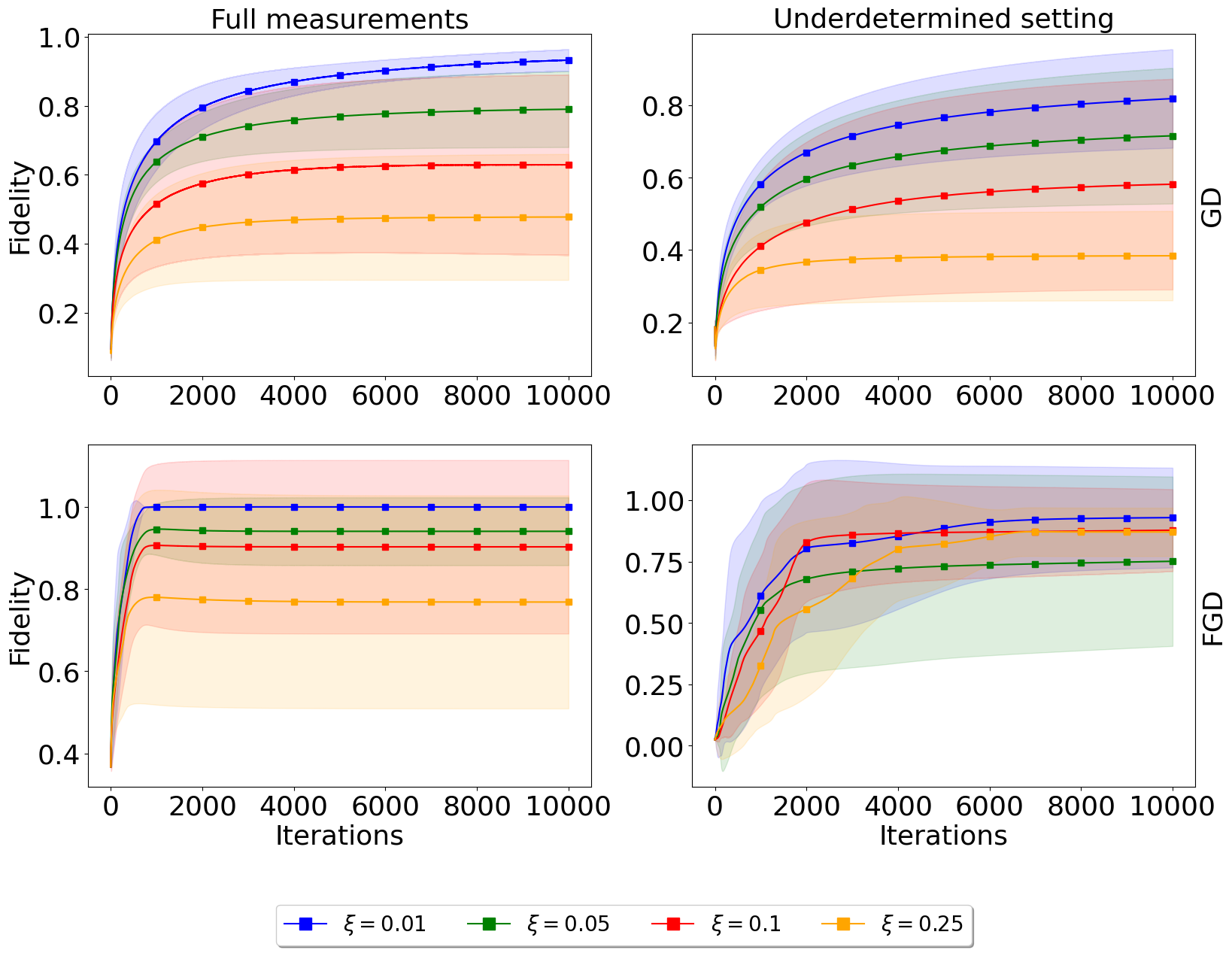}
    \caption{2 qubit Gaussian noise. The top and the bottom figures show results from the \texttt{adaGD} and the \texttt{adaFGD} algorithms, respectively. The left and right figures show results on 128 measurements and 96 measurements, respectively.}
    \label{fig:2_qubit_gaussian_fidelity_paper}
\end{figure}

\medskip
\noindent \textbf{Gaussian noise.} In Figure \ref{fig:2_qubit_gaussian_fidelity_paper}, the full measurements setting showed improvements when Gaussian noise was added, although with slightly higher variance. We attribute such a behavior to the uncertainty of noise added in the measurement stage, since this leads to an external element to the process $\chi$ being applied, and thus a biased yet consistent $\hat{\chi}$ will be obtained. Despite this, we also find a drastic speedup in convergence. For the underdetermined setting, \texttt{FGD} consistently obtained better fidelity than GD for all values of $\xi$, while still showing great variance for $\xi=0.01$. While \texttt{FGD} in this setting does not converge as quickly, it is to be expected due to the measurement defficiency.

\begin{figure}[h]
    \centering
    \includegraphics[width=0.5\textwidth]{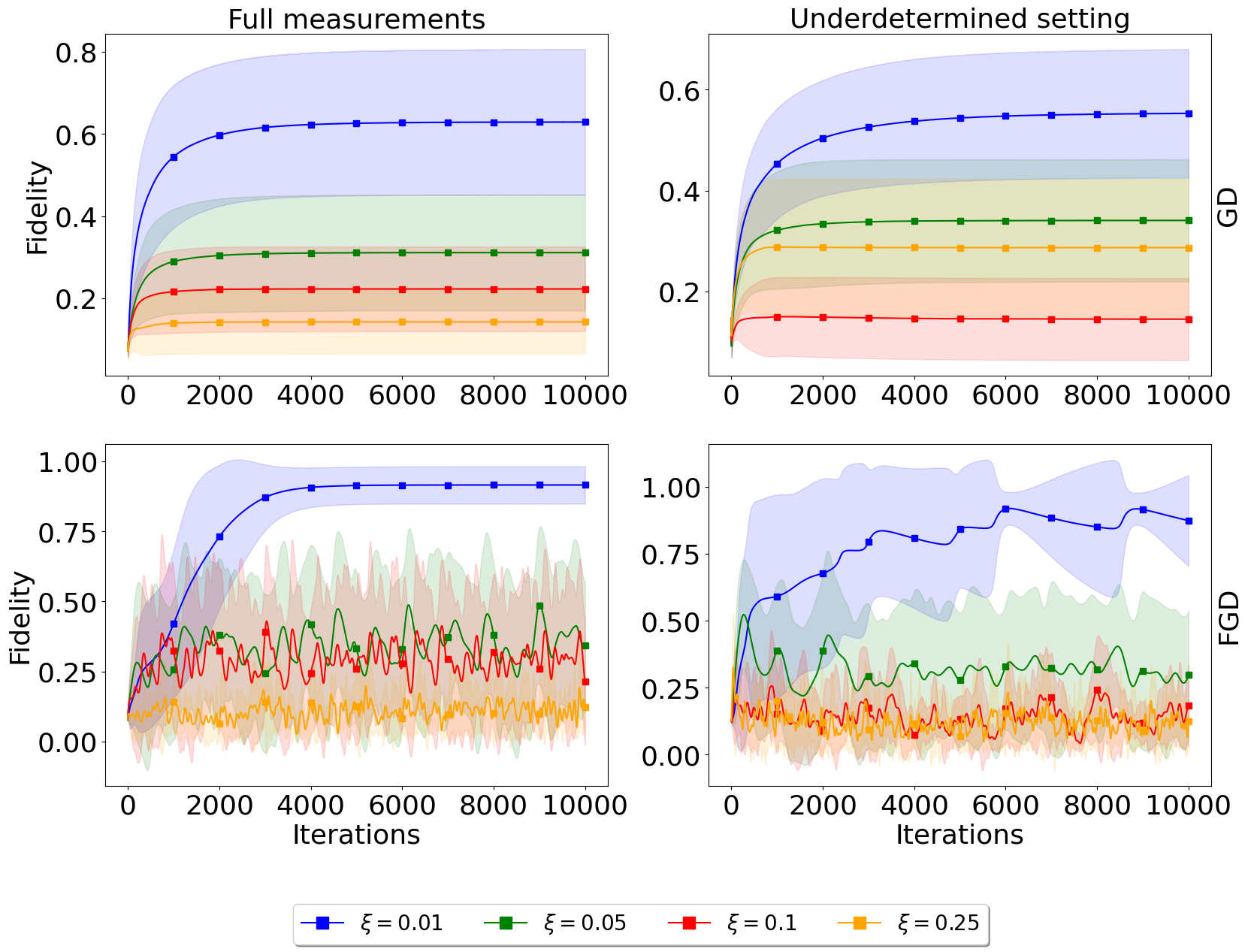}
    \caption{2 qubit coherent noise. The top and the bottom figures show results from the \texttt{adaGD} and the \texttt{adaFGD} algorithms, respectively. The left and right figures show results on 128 measurements and 96 measurements, respectively.}
    \label{fig:2_qubit_coherent_fidelity_paper}
\end{figure}

\medskip
\noindent \textbf{Coherent noise.} From Figure \ref{fig:2_qubit_coherent_fidelity_paper} we can observe that coherent noise effectively reduces fidelity to a great extent, as is explained in the literature. For GD, this reduction is consistent starting from $\xi = 0.01$ on both measurement settings. \texttt{FGD} shows an unstable behavior for coherent noise on $\xi > 0.01$, although with a slightly larger fidelity in the case of the full measurement setting. For the same setting, we obtained an average fidelity surprisingly near the optimal fidelity for $\xi = 0.01$. This occurrence hints towards better estimates of $\hat{\chi}$ with smaller values of $\xi$ when coherent noise is applied. For the underdetermined setting, on the other hand, the fidelity on all values of $\xi$ were inconsistent, although $\xi=0.01$ obtains great improvements in average fidelity than its GD counterpart, with some runs obtaining an optimal $\chi^*$. Starting from $\xi=0.05$, coherent noise yields deficient fidelity in this setting.

\begin{figure}[h]
    \centering
    \includegraphics[width=0.5\textwidth]{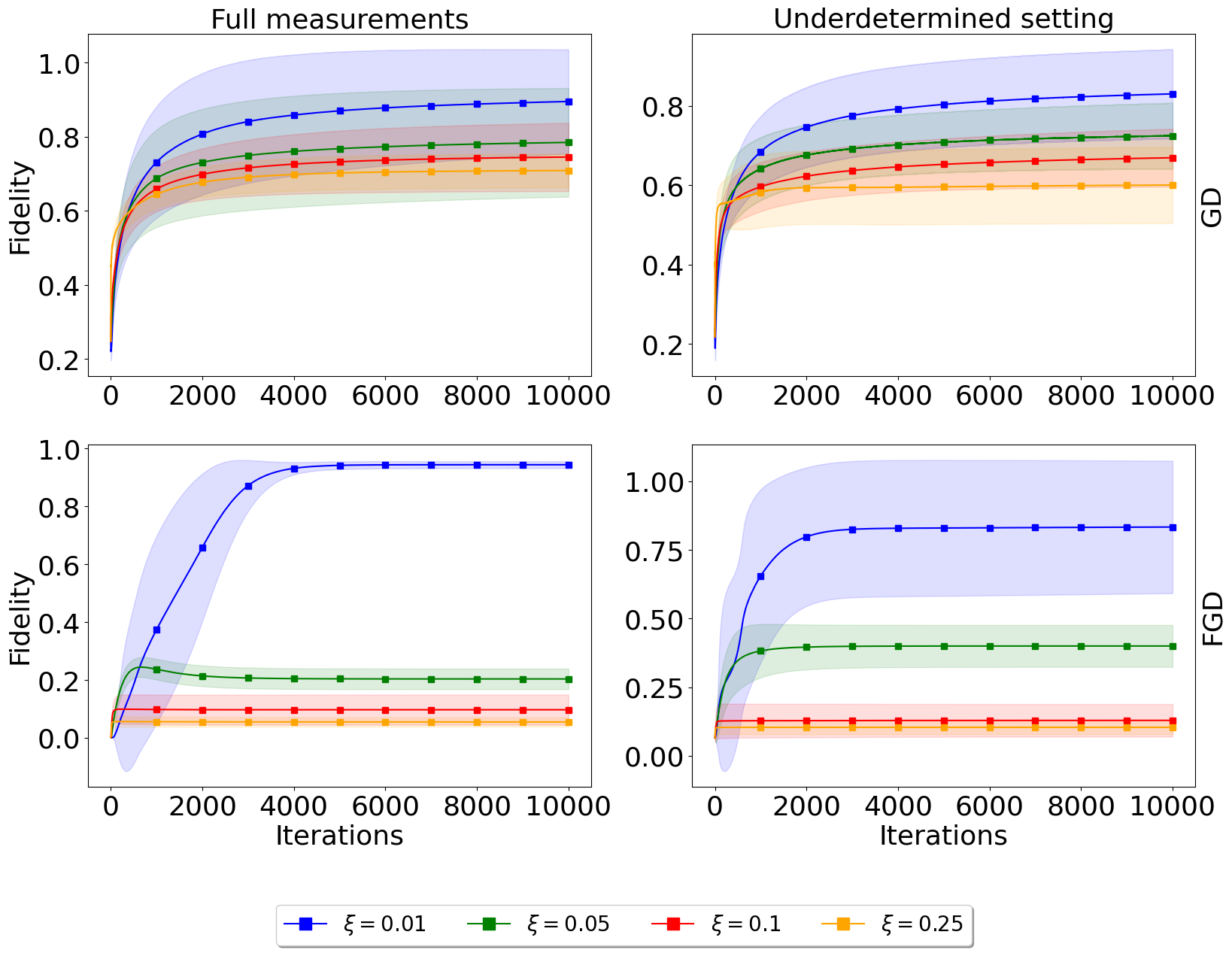}
    \caption{2 qubit incoherent noise. The top and the bottom figures show results from the \texttt{adaGD} and the \texttt{adaFGD} algorithms, respectively. The left and right figures show results on 128 measurements and 96 measurements, respectively.}
    \label{fig:2_qubit_incoherent_fidelity_paper}
\end{figure}

\medskip
\noindent \textbf{Incoherent noise.} For the results of Figure \ref{fig:2_qubit_incoherent_fidelity_paper} on our general incoherent noise model, all $\xi > 0.01$ values tested in both measurement settings yielded low fidelities for \texttt{FGD} compared to the same values on GD. For the case of $\xi=0.01$, however, we obtained the optimal $\chi^*$ on the full measurement setting, and a suboptimal $\hat{\chi}$ for the underdetermined setting, with the same average fidelity on both \texttt{FGD} and GD. The main difference in these two sets of results comes from some runs on \texttt{FGD} retrieving $\hat{\chi} = \chi^*$ and thus obtaining very high fidelities in some runs. We observe the same trend of rapid convergence when using this noise model.

\texttt{FGD} reached convergence on all models tested except for the underdetermined setting of the coherent noise model, where $f(\chi)$, $\| \hat{\chi} - \chi^* \|$, and $F(\hat{\chi}, \chi^*)$ did not converge but still managed to reach a better estimate for $\hat{\chi}$. On average, good performance on \texttt{FGD} was obtained for reasonable noise, and convergence was reached much quicker (about $5\times$) on average, except for the case where coherent noise is applied in the underdetermined setting. Contrary to GD, \texttt{FGD} always reached a near-optimal $\hat{\chi}$ for small noise levels when the complete set of measurements was available, and for both settings except on the general incoherent noise model. Despite this, representing incoherent noise as depolarizing noise may provide better guarantees.

    \label{fig:third}

\section{Conclusion}
{\label{sec:conclusions}}

As our concluding remarks, we applied the non-convex factored gradient descent algorithm to QPT and studied critical aspects such as the number of measurement settings in both a scenario where full measurements are available for our selection of initial states and measurement operators, and an underdetermined setting where not all measurements are available, aligning with a compressed sensing paradigm. We observed how a maximum of $2 \cdot 8^n$ circuit configurations could completely characterize a process matrix $\chi$, and $\mathcal{O}(4^n)$ measurements were tested, yielding great improvements in fidelity using FGD. This shows a substantial improvement in the current configurations for out-of-the-box QPT solutions, and provides an empirical loose lower bound on circuit configurations such that QPT would remain easy to prepare yet more effective and with better scaling on the number of qubits. We compared \texttt{FGD} with GD in order to understand which noise models led to better performance for FGD. Our results indicate that \texttt{FGD} performs best on depolarizing noise models and Gaussian noise models, along with an inconsistent behavior in the case of coherent noise, although potentially better at estimating a process affected by coherent noise than GD is. While \texttt{FGD} was unable to accurately estimate a quantum process when a general model for incoherent noise with large levels of noise was applied, using the depolarizing noise model may provide a good-enough generalization of incoherent noise, and much better results. Future work is to provide theory for the \texttt{FGD} algorithm applied to QPT, and a complete study on measurement reduction such as using the minimal initial state set of \cite{baldwin_quantum_2014} alongside the $2d$ POVMs used in this study for a minimal number of total circuit configurations. Another goal for the \texttt{FGD} algorithm applied to QPT is to determine the least number of measurement settings necessary to obtain good fidelity measures, in both the fully determined case and in the underdetermined case.

\bibliography{QPT}
\bibliographystyle{IEEEtran}

\end{document}